\documentclass[12pt]{article}

\usepackage{scicite}

\usepackage{times}

\usepackage{graphicx}
\usepackage{nicefrac} 
\usepackage[hidelinks]{hyperref} 
\usepackage{graphicx}
\usepackage{color}
\usepackage{siunitx} 
\DeclareSIUnit\Angs{\angstrom}

\newcommand{\tise}{TiSe$_2$~}
\newcommand{\ttise}{TiSe$_2$}

\newcommand{\flux}{\,\nicefrac{$\mu J$}{$cm^2$}\,}
\newcommand{\kp}{\si{\Angs}$^{\text{-}1}$}
\newcommand{\kpp}{$k_\parallel$}

\newenvironment{sciabstract}{%
\begin{quote} \bf}
{\end{quote}}

\title{Mapping the dispersion of the occupied and unoccupied band structure in photoexcited 1T-\tise} 

\author{Maximilian Huber $^{1}$, Yi Lin $^{1}$, Nicholas Dale$^{1,2}$, Renee Sailus$^{3}$, Sefaattin Tongay$^{3}$, \\ Robert A.\ Kaindl $^{1,4}$ and Alessandra Lanzara$^{1,2\ast}$\\
\\
\normalsize{$^{1}$Materials Science Division, Lawrence Berkeley National Laboratory,}\\
\normalsize{Berkeley, CA 94720, USA}\\
\normalsize{$^{2}$Physics Department, University of California Berkeley, Berkeley, CA 94720, USA}\\
\normalsize{$^{3}$ Materials Science and Engineering Department, Arizona State University}\\
\normalsize{AZ 85281, USA}\\
\normalsize{$^{4}$ Department of Physics and CXFEL Labs, Arizona State University, AZ 85287, USA}\\
\normalsize{$^\ast$To whom correspondence should be addressed; E-mail:  alanzara@lbl.gov}
}

\date{}

\begin{document} 


\baselineskip24pt


\maketitle

\begin{sciabstract}
Charge density waves (CDW) are states of broken symmetry with a periodic modulation of charge and lattice typically leading to the opening of a gap in the band structure. In the model CDW system 1T-\tise such a gap opens up between its Se$_{4p}$ valence and Ti$_{3d}$ conduction band, accompanied by a change of dispersion. These changes are crucial in understanding the CDW phase, as they provide a measure of the Se$_{4p}$-Ti$_{3d}$ hybridization strength and characteristic mechanistic features. Using time- and angle-resolved photoelectron spectroscopy (trARPES), the unoccupied band structure is populated with near-infrared (NIR) pump pulses which allows to to directly visualize the parabolically-shaped Ti$_{3d}$ conduction band. Furthermore, we observe a transient change of effective mass in the Se$_{4p}$ valence band following photoexcitation. This occurs alongside an overall reduction due to weakening of the CDW phase and is accompanied by an oscillating component with the frequency of the characteristic A$_{1g}$ phonon. These observations, enabled by trAPRES, highlight the importance of the lattice contributions in establishing the CDW order in \mbox{1T-\ttise}.

\end{sciabstract}


\section*{Introduction}

The question of the origin of the charge-density wave (CDW) state in 1T-\tise remains up to now puzzling and is topic of recent research. It is commonly accepted that the CDW phase is stabilized by a combination of the pseudo Jahn-Teller effect and excitonic insulator mechanism\,\cite{Porer2014a}, however the exact interplay of both effects remains elusive. Equilibrium studies have mostly suggested a pseudo Jahn-Teller effect~\cite{Rossnagel2010, Rossnagel2002, Rossnagel2011, Wegner2018}, strongly supported by the observed phonon softening\,\cite{Holt2001, Weber2011} although an excitonic mechanism has also been proposed, based on the report of plasmon softening\,\cite{Kogar2017}. The latter scenario is further supported by time-resolved studies, where an ultrafast melting of CDW order\,\cite{Rohwer2011, Hellmann2012, Mathias2016} at high excitation densities has been reported, whose timescale and fluence scaling are suggestive of an electronically mediated CDW phase. \\ 
Angle resolved photoemission spectroscopy (ARPES) is the ideal tool to study charge density wave order and many-body interactions, as it provides direct access to the quasiparticle band structure. The combination with a pump beam in a time resolved ARPES (trARPES) setup allows to populate in equilibrium unoccupied states and study photoinduced dynamics of excited carriers and scattering processes\,\cite{Smallwood2012x, Mathias2016, HuberX}. \\ 
In this work we use extreme-ultraviolet (XUV) trARPES\,\cite{Buss2019} to study the dispersion of the band structure above the Fermi level as well as the Se$_{4p}$ valence band at $\Gamma$ point. We find that optical excitation populates a second parabolically-shaped branch of the conduction band which we associate with the Ti$_{3d}$ band. Additionally, at the $\Gamma$ point, we reveal oscillations of the Se$_{4p}$ valence band position at a frequency comparable with the A$_{1g}$ phonon characteristic of the CDW phase. These oscillations can be only seen at the top of the valence band, hinting towards a transient change of dispersion and effective mass renormalization, thus exemplifying the importance that Jahn-Teller interactions also play in establishing CDW order.

\section*{Results and Discussion}

\begin{figure} [h]
	\centering
	\includegraphics[width=0.85\textwidth]{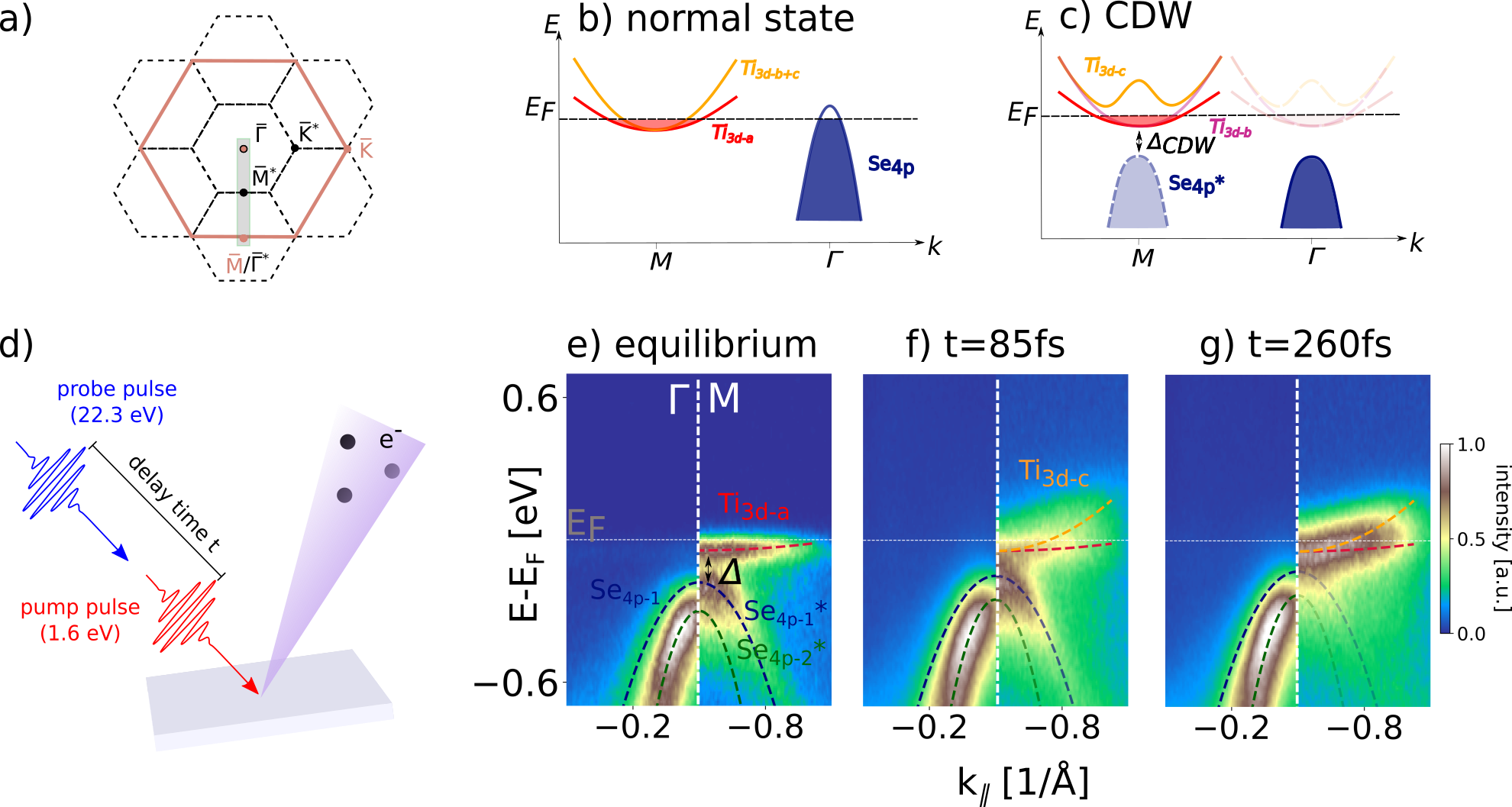}
	\caption{ a) First BZ in the high symmetry (orange) and CDW phase (black); the grey bar illustrates the cuts measured in the ARPES spectra in $\Gamma$-$M$ direction. b-c) Schematic band structure\,\cite{Monney2009} of the (b) high symmetry as well as (c) CDW state. Note that in both phases three Ti$_{3d}$ bands are present with their center being threefold degenerate in the high symmetry phase. d) Schematic representation of a trARPES experiment. Pump pulses with 1.6\,eV excite the sample, which is subsequently probed by 22.3\,eV XUV probe pulses. e-g) ARPES spectra of the $\Gamma$ and $M$-point at 80\,K in equilibrium\,(e) and after excitation with 780\,nm pump pulses at 80\,\flux after 85\,fs\,(f) and after 260\,fs\,(g).  The Fermi level is indicated by the dashed gray line.}
	\label{fig:fig1}
	\end{figure}

\autoref{fig:fig1} shows the dynamics of the electronic band structure at the $M$ and $\Gamma$ point upon pump excitation along the high symmetry direction, $\Gamma-M$ (see \autoref{fig:fig1}a). Data are taken in the CDW state, for T=\,80\,K and at a fluence of 80\,\flux. \autoref{fig:fig1}a shows the 2D Brillouin zone (BZ) in the high symmetry (orange) and CDW (black) state. When going through the phase transition, the unit cell size doubles from a (1$\times$1$\times$1) structure at room temperature to a (2$\times$2$\times$2) superstructure below $\sim$200\,K\,\cite{DiSalvo1976}. Figures\,1b and\,c show a schematic of the calculated band structure -adapted from\,\cite{Monney2009}- corresponding respectively to the high symmetry and CDW state at $\Gamma$ and $M$ point with a focus on the Ti$_{3d}$ conduction and the Se$_{4p}$ valence band. We note that the Se$_{4p}$ band, derived from  Se${4p_{x,y}}$\,\cite{Rossnagel2002} orbitals, is spin-orbit split into two individual bands, Se$_{4p\text{-}1}$ and Se$_{4p\text{-}2}$, which was neglected in this calculation. Fingerprints of the CDW state are the backfolding of the Se$_{4p}$ bands onto the $M$ point Se$_{4p}$* and the opening of a gap\,\cite{Rossnagel2002, Rossnagel2010, Rossnagel2011, Rohwer2011, Kidd2002}. While the gap is formed between the lowest conduction band, the Ti$_{3d\text{-}a}$ band, and the upper Se$_{4p\text{-}1}$* band, the former does not interact with the Se band and remains unchanged in both the high and low temperature phase\,\cite{Monney2009, Watson2019}. Instead it is the Ti$_{3d\text{-}c}$ band that hybridizes with the center of the Se$_{4p}$ band\,\cite{Hellgren2017, Chen2018}. 
A schematic representation of our experiment is given in \autoref{fig:fig1}d. NIR Pump pulses with 780\,nm ($\sim$1.6\,eV) excite the sample at 80\,K, which is then subsequently probed by 22.3\,eV XUV probe pulses. The angle and energy of the photoemitted electrons are detected with a hemispherical analyzer. Details to the setup can be found in\,\cite{Buss2019}. Raw energy vs momentum maps are shown in \autoref{fig:fig1}e-g at equilibrium and for different delay times.  At equilibrium (\autoref{fig:fig1}e), although both spin-orbit split Se$_{4p}$ bands are resolved, matrix element effects allow us to visualize them one at a time. Specifically at the $M$ point we can clearly resolve the Se$_{4p\text{-}1}$ band while the Se$_{4p\text{-}2}$ is better visible at the $\Gamma$ point. We can additionally resolve the bottom of the Ti$_{3d\text{-}a}$  conduction band at the $M$  point and hence access the CDW gap ($\Delta$). A number of clearly visible changes in the excited state (Figures\,1f and\,g) can be observed in agreement with earlier work\,\cite{HuberX, Rohwer2011, Mathias2016, Buss2019}. This includes (i) population of states above the Fermi level by hot electrons, in particular the occupancy of the previously unoccupied parabolic band, which we assign tentatively to Ti$_{3d\text{-}c}$ (compare \autoref{fig:fig1}b and\,c); (ii) a partial closing of the gap between the Ti$_{3d\text{-}a}$  and Se$_{4p\text{-}1}$ states (\autoref{fig:fig1}f) at $M$, accompanied by (iii) an upshift of the Se$_{4p\text{-}2}$ band at $\Gamma$ and (iv) the  disappearance of the Se$_{4p\text{-}1}$ backfolded band at $M$ in concomitance with the gap closing at later delay time ($t=$260\,fs,\,\autoref{fig:fig1}g).

\begin{figure} [h]
	\centering
	\includegraphics[width=0.7\textwidth]{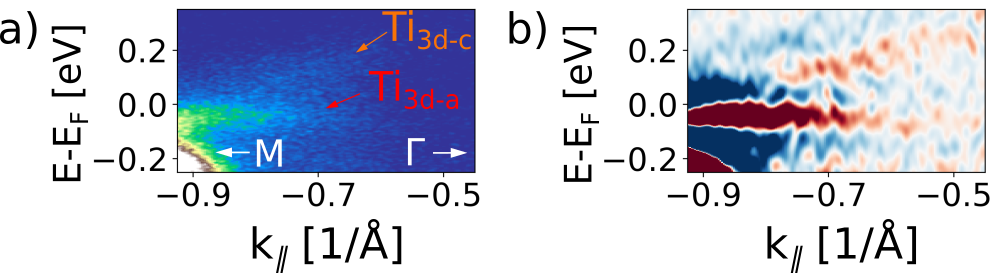}
	\caption{a) ARPES image plot after excitation with 70\,\flux after 100\,fs  b) Corresponding second derivative image after smoothing with the Gaussian method over a window of 30\,meV. }
	\label{fig:fig2}
	\end{figure}
 
To study the excited states in more detail, we present in \autoref{fig:fig2} ARPES maps near the Fermi level which were averaged over extened measurement times. The photoexcited data at 100\,fs is shown in \autoref{fig:fig2}a alongside with its second derivative in \autoref{fig:fig2}b. 
These data reveal that a significant amount of hot electrons occupy states above the Fermi level in what resembles an upwards-facing -electron-like- parabola in accordance with \autoref{fig:fig1}e-g. This newly populated orbital is different from the less-dispersing, in equilibrium partially occupied Ti$_{3d\text{-}a}$ orbital. Moreover, the expectation that the Ti$_{3d\text{-}a}$ orbital should not considerably change its shape\,\cite{Monney2009, Bianco2015, Watson2019}, as it barely hybridizes with the Se$_{4p}$ valence band or shift upwards during the phase transition to the high temperature state\,\cite{Monney2010a} implies that another orbital which is distinct from Ti$_{3d\text{-}a}$ gets transiently populated. This cannot be the Ti$_{3d\text{-}b}$ band which should carry negligible spectral weight\,\cite{Monney2009}. Therefore we assign this band to the Ti$_{3d\text{-}c}$ band\,\cite{Monney2009, Watson2019}, which hybridizes with the Se valence band and is thus sensitive to CDW formation. From \autoref{fig:fig2}a it appears like the Ti$_{3d\text{-}a}$ and Ti$_{3d\text{-}c}$ bands are merging into each other at $M$  point, however we note that for this pump fluence the CDW phase gets strongly perturbed, i.e.\ the splitting between the bands is significantly reduced and is thus most likely simply too small to be resolved in our experiment.\\ \\
Our ability to resolve this band constitutes the first step toward a more complete understanding of the mechanism behind the CDW transition in \ttise. Indeed, in a Jahn-Teller scenario the degeneracy of the bottom of the parabolic-like Ti$_{3d\text{-}a}$ and Ti$_{3d\text{-}c}$ is lifted\,\cite{Rossnagel2010, Watson2019}, leading to an upshift of the hybridising Ti$_{3d\text{-}c}$ center. On the other hand, within an excitonic insulator scenario more pronounced changes of the band structure are expected\,\cite{Monney2009, Cazzaniga2012}, with the presence of a characteristic double-well shape predicted by Kohn in his seminal work\,\cite{Kohn1967}. In general, however, it is a very delicate task to observe bands above the Fermi level corresponding to the CDW state experimentally, since the commonly used 780\,nm pump wavelength mostly excites electrons high above the Fermi level\,\cite{Rohde2014, HuberX}. Thus, by the time these carriers scatter down to lower energies, the CDW phase is already strongly perturbed or quenched, thus displaying the high symmetry band structure. Further studies using high resolution XUV-trARPES systems with MIR pump could potentially solve these problems, driving a direct excitation into the Ti$_{3d\text{-}c}$ orbital\,\cite{Monney2016} which could provide clear evidence for a Mexican hat shaped conduction band.

\begin{figure} [h]
	\centering
	\includegraphics[width=0.8\textwidth]{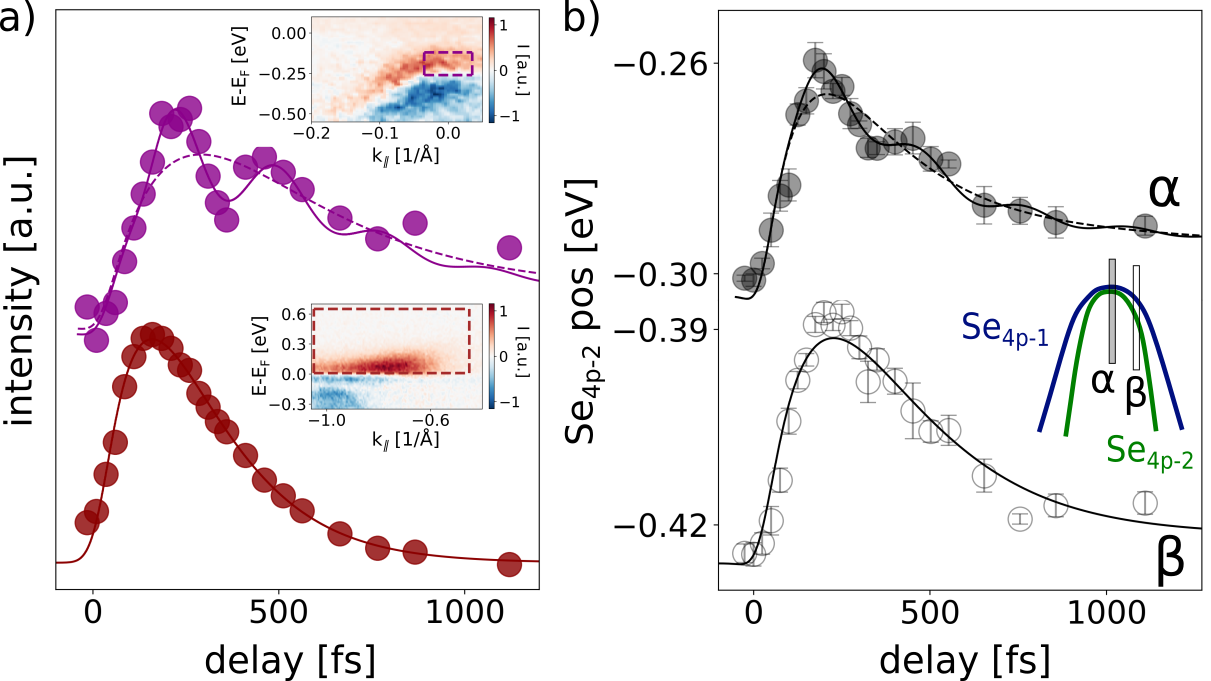}
	\caption{ a) Dynamics of the intensity integrated above the Se$_{4p\text{-}2}$ band (magenta, integration region shown in the inset) versus the dynamics of excited carriers (red, integration region see inset) after excitation with 80\,\flux. Solid lines are fits to the data. b) Momentum resolved dynamics of the Se$_{4p\text{-}2}$ position after excitation with 80\,\flux. Solid lines are fits to the data. Curve $\alpha$ was obtained by fitting EDCs taken right at $\Gamma$ point, curve $\beta$ from EDCs taken at \kpp=-0.1\kp. The integration regions for the EDCs is schematically depicted in the inset.}
	\label{fig:fig3}
	\end{figure}

We now focus our attention to the dynamical evolution of the Se$_{4p}$ valence band following photo-excitation. As the intensity of this band at $\Gamma$ point stays constant over all delays, integrating slightly above the position of the band maximum in equilibrium can give information about transient band shifts following optical excitation.  In \autoref{fig:fig3}a we show this integrated spectral weight, with the integration region indicated in the difference spectrum in the inset. In addition to the sudden increase of intensity following photoexcitation, the data also reveal an oscillating component as a function of delay time.
To best describe the dynamical evolution of the data over time we thus use a fitting function consisting of three exponentials (one describing the rise time and a fast and slow component for the recovery time\,\cite{HuberX}) as well as a damped cosine function convoluted with a Gaussian accounting for the experimental time resolution. We find an oscillation frequency of 3.48 $\pm 0.24$\,THz, which is comparable to the A$_{1g}$ phonon mode characteristic of the CDW state\,\cite{Snow2003}. In contrast, the intensity of the excited carriers shows only a sudden increase followed by a fast decrease without any sign of oscillations. \\
In \autoref{fig:fig3}b we study the time delay dependence of the valence band energy position. To extract the energy vs momentum position of the valence band, EDCs were fitted using two peaks 
of Voigt lineshape on a linear background (accounting for the Se$_{4p\text{-}2}$ and a Se$_{4p\text{-}z}$\,\cite{Cazzaniga2012} band). The position of the Se$_{4p\text{-}2}$ peak position at the band maximum at \kpp=0\kp(i.e.\ $\Gamma$ point, denoted as $\alpha$), reveal an overall upshift of the band followed by a downshift to slightly above the equilibrium position at later delay times. The upshift of the valence band is indicative of a partial closing of the gap, suggesting a weakening of the CDW phase. A closer look reveals that the valence band position also shows an oscillating component. The frequency of the oscillation is similar to the one observed for the Se$_{4p}$ intensity, with frequencies of 3.40 $\pm 0.47$ THz and 3.48 $\pm 0.24$\,THz, comparable to the phonon mode frequency and thus implying a strong intrinsic coupling between the atomic motion and the electronic structure\,\cite{Monney2016}. To better understand the changes in the band structure due to the oscillations, in \autoref{fig:fig3}b we show the Se$_{4p\text{-}2}$ peak position away from the maximum of the Se$_{4p}$ band at \kpp=-0.1\,\kp (denoted as $\beta$) as a function of the delay time. Interestingly, the position shows an overall upward shift towards the Fermi level similar to the one observed at $\alpha$, reaching its maximum at the same time and then slowly decaying back toward equilibrium without any oscillation, in contrast to the behavior of the band maximum position $\alpha$. The different response of the top ($\alpha$) and side ($\beta$) of the Se$_{4p}$ band to photoexcitation suggests a transient change of effective mass, probably with this phonon frequency.

\clearpage

\begin{figure} [h]
	\centering
	\includegraphics[width=0.8\textwidth]{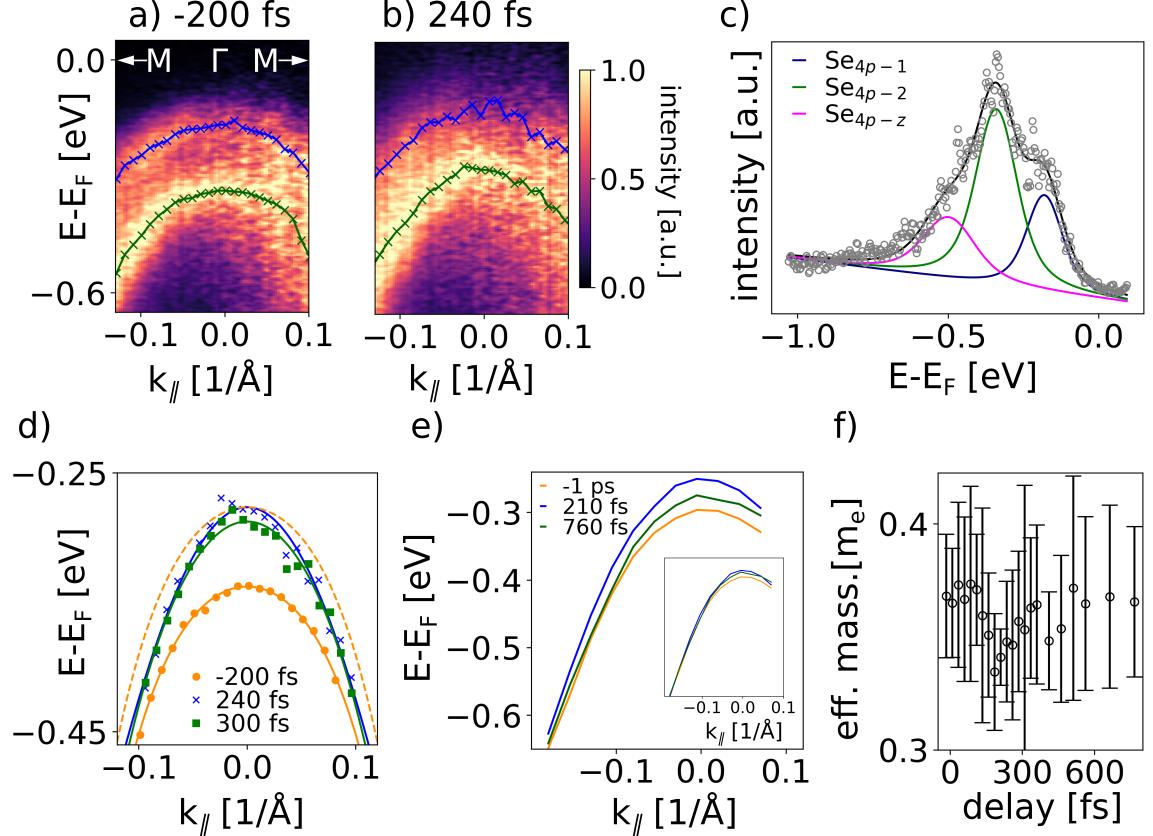}
	\caption{ a-b) ARPES maps around the $\Gamma$-point (a) before excitation  and (b) 240\,fs after excitation with 90\,\flux pump fluence. In order to compensate for matrix element effects spectra are normalized so that every EDC slice carries the same maximum and minimum intensity. The fitted band position for the Se$_{4p\text{-}1}$ and Se$_{4p\text{-}2}$ band are shown in blue and green, respectively. c) EDC for a time delay of t=-200\,fs at momentum \kpp=0\kp along with fits. d) Se$_{4p\text{-}2}$ band dispersion extracted from fitting EDCs at t=-200\,fs (orange), 240\,fs (blue) and 300\,fs (green) after excitation with 90\flux. Solid lines are fits to the data using a fourth order polynomial. For a direct comparison a copy of the orange line is shifted upward in energy. e) Se$_{4p\text{-}2}$ band dispersion extracted from fitting EDCs at t=-1\,ps (orange), 210\,fs (blue) and 760\,fs (green) after excitation with 80\flux fluence. The inset shows the same bands but shifted in energy so that the band positions at -0.15\kp are aligned. f) Effective mass of the Se$_{4p\text{-}2}$ band after excitation with 80\flux extracted from parabolic fits to the dispersion.}
	\label{fig:fig4}
	\end{figure}


Motivated by these results, in \autoref{fig:fig4} we analyze the dispersion of the selenium valence band in more detail to directly extract information on the effective mass.  
\autoref{fig:fig4}a-b shows  the transient ARPES map around $\Gamma$ before excitation (t=-200\,fs) and for t=240\,fs after excitation with a fluence of 90\flux, respectively. For this sample, both spin-orbit split Se bands can be clearly identified. This is clear in the EDCs (\autoref{fig:fig4}c) where one can easily distinguish the two peaks at about -0.18\,eV, -0.32\,eV as well as a shoulder at -0.5\,eV, which are, respectively, attributed to the  Se$_{4p\text{-}1}$, Se$_{4p\text{-}2}$ and Se$_{4p\text{-}z}$ orbitals\,\cite{Cazzaniga2012}. To enhance clarity and compensate for matrix-element effects we normalize the intensity of the spectrum for each energy position. The fitted band positions of both Se$_{4p\text{-}1}$ and Se$_{4p\text{-}2}$ band over momentum space are superimposed in blue and green markers, respectively. One can clearly see that  both bands shift significantly up in energy after optical excitation while simultaneously becoming more parabolic, indicative of a decrease of effective mass. To make the change in band dispersion clearer, in \autoref{fig:fig4}d plots the extracted band dispersion of the Se$_{4p\text{-}2}$ valence band before excitation and in the excited state at two different delays. To describe the dispersion over a broad momentum range, we fit the data points with a fourth-order polynomial. Between -200\,fs (orange data points) and 240\,fs (blue data points) the whole band shifts up about 55\,meV, accompanied by a clear change of effective mass. For a direct comparison we shifted the fitted dispersion of the equilibrium band dispersion up to the position of the excited band (dotted orange line). Interestingly, between 240 and 300\,fs the center of the Se$_{4p\text{-}2}$ band (\kpp $\sim$ -0.07 to 0.07) is shifted slightly downward, while the edges of the band are almost at the same position, resulting in an increase of effective mass. This indicates that the changes of effective mass at the band center are decoupled form the overall band shift. 
Having established this clear change in effective mass, we analyze the dispersion of the Se valence band from the dataset shown in Figures\,1 and\,3 in more detail. \autoref{fig:fig4}e shows the extracted dispersion at three different time delays, shown over a larger momentum range than in \autoref{fig:fig4}d. The inset shows the same plots shifted in energy so that the band edges at -0.15\kp are aligned, showing again how the top of the band becomes modulated in a different way than the band edges away form the center, thus further reproducing the results of \autoref{fig:fig4}d. In \autoref{fig:fig4}f we show the dynamical evolution of the effective mass, extracted from parabolic fits to the band dispersion between -0.15 and 0.08\,\kp. A systematic decrease of the effective mass followed by a partial recovery can be observed. \\ \\

The decrease of effective mass corresponding to the perturbation of the CDW phase is consistent with both the screening of excitons in an excitonic insulator scenario\,\cite{Cazzaniga2012, Monney2009} and a weakening of the hybridization between Se$_{4p}$ and Ti$_{3d}$  states in a Jahn-Teller scenario\,\cite{Watson2019, Rossnagel2010, Rossnagel2011}. On the other hand, the oscillations of the Se$_{4p}$ band at the A$_{1g}$ phonon frequency are an unique signature of the Jahn-Teller effect. Indeed, within an excitonic insulator scenario the dynamics of the excited carriers should be similar to the dynamics of the band structure. The absence of oscillation on the former support the role of the lattice in defining the Se$_{4p}$ band shape. Therefore, the observed flattening of the Se$_{4p}$ band when going from high temperature to CDW phase is very likely connected to the lattice. Nevertheless, while our results point towards the important role the Jahn-Teller effect plays in establishing CDW order they do not exclude that strong electronic correlations are present as well. Indeed, in theoretical work it is predominantly proposed that a cooperative mechanism between both concepts is likely responsible for the formation of the low temperature CDW phase\,\cite{VanWezel2010, Kaneko2018, Lian2020}.

\section*{Conclusion}
In conclusion, the results here presented show that the plethora of effects observed in \tise provides a unique platform for studying the intriguing interplay of many-body couplings in a strongly correlated system. In this regard we observed how the effective mass, indicative of the Se$_{4p}$-Ti$_{3d}$ hybridization strength decreases after optical excitation while the dispersion is modulated by phonon oscillations. This emphasizes the role that Jahn-Teller interactions play in the establishment of CDW order in \ttise. Finally, we were also able to resolve sides of the Ti$_{3d\text{-}c}$ band above the Fermi level.

\section*{Data Availability} 
The datasets generated and/or analyzed during the current study are not publicly available as they contain additional findings not reported in this manuscript,  but are available from the corresponding author upon reasonable request.

\clearpage




\bibliographystyle{IEEEtran}
\bibliography{TiSe16}

\begin{thebibliography}{10}
\providecommand{\url}[1]{#1}
\csname url@samestyle\endcsname
\providecommand{\newblock}{\relax}
\providecommand{\bibinfo}[2]{#2}
\providecommand{\BIBentrySTDinterwordspacing}{\spaceskip=0pt\relax}
\providecommand{\BIBentryALTinterwordstretchfactor}{4}
\providecommand{\BIBentryALTinterwordspacing}{\spaceskip=\fontdimen2\font plus
\BIBentryALTinterwordstretchfactor\fontdimen3\font minus
  \fontdimen4\font\relax}
\providecommand{\BIBforeignlanguage}[2]{{%
\expandafter\ifx\csname l@#1\endcsname\relax
\typeout{** WARNING: IEEEtran.bst: No hyphenation pattern has been}%
\typeout{** loaded for the language `#1'. Using the pattern for}%
\typeout{** the default language instead.}%
\else
\language=\csname l@#1\endcsname
\fi
#2}}
\providecommand{\BIBdecl}{\relax}
\BIBdecl

\bibitem{Porer2014a}
M.~Porer, U.~Leierseder, J.~M. M{\'{e}}nard, H.~Dachraoui, L.~Mouchliadis,
  I.~E. Perakis, U.~Heinzmann, J.~Demsar, K.~Rossnagel, and R.~Huber,
  ``{Non-thermal separation of electronic and structural orders in a persisting
  charge density wave},'' \emph{Nature Materials}, vol.~13, no.~9, pp.
  857--861, 2014.

\bibitem{Rossnagel2010}
K.~Rossnagel, ``{Suppression and emergence of charge-density waves at the
  surfaces of layered 1T$-$TiSe$_2$ and 1T$-$TaS$_2$ by in situ Rb
  deposition},'' \emph{New Journal of Physics}, vol.~12, 2010.

\bibitem{Rossnagel2002}
K.~Rossnagel, L.~Kipp, and M.~Skibowski, ``{Charge-density-wave phase
  transition in (formula presented): Excitonic insulator versus band-type
  Jahn-Teller mechanism},'' \emph{Physical Review B - Condensed Matter and
  Materials Physics}, vol.~65, no.~23, pp. 1--7, 2002.

\bibitem{Rossnagel2011}
K.~Rossnagel, ``{On the origin of charge-density waves in select layered
  transition-metal dichalcogenides},'' \emph{Journal of Physics Condensed
  Matter}, vol.~23, no.~21, 2011.

\bibitem{Wegner2018}
A.~Wegner, J.~Zhao, J.~Li, J.~Yang, A.~A. Anikin, G.~Karapetrov, D.~Louca, and
  U.~Chatterjee, ``{Evidence for breathing-type pseudo Jahn-Teller distortions
  in the charge density wave phase of 1T-TiSe$_2$},'' \emph{Physical Review B},
  vol. 101, pp. 1--6, 2020.

\bibitem{Holt2001}
M.~Holt, P.~Zschack, H.~Hong, M.~Y. Chou, and T.~C. Chiang, ``{X-ray studies of
  phonon softening in TiSe$_2$},'' \emph{Physical Review Letters}, vol.~86,
  no.~17, pp. 3799--3802, 2001.

\bibitem{Weber2011}
F.~Weber, S.~Rosenkranz, J.~P. Castellan, R.~Osborn, G.~Karapetrov, R.~Hott,
  R.~Heid, K.~P. Bohnen, and A.~Alatas, ``{Electron-phonon coupling and the
  soft phonon mode in TiSe$_2$},'' \emph{Physical Review Letters}, vol. 107,
  no.~26, pp. 1--5, 2011.

\bibitem{Kogar2017}
A.~Kogar, M.~S. Rak, S.~Vig, A.~A. Husain, F.~Flicker, Y.~I. Joe, L.~Venema,
  G.~J. Macdougall, T.~C. Chiang, E.~Fradkin, J.~V. Wezel, and P.~Abbamonte,
  ``{Signatures of exciton condensation in a transition metal
  dichalcogenide},'' \emph{Science}, vol. 1317, no. December, pp. 1314--1317,
  2017.

\bibitem{Rohwer2011}
T.~Rohwer, S.~Hellmann, M.~Wiesenmayer, C.~Sohrt, A.~Stange, B.~Slomski,
  A.~Carr, Y.~Liu, L.~M. Avila, M.~Kall{\"{a}}signne, S.~Mathias, L.~Kipp,
  K.~Rossnagel, and M.~Bauer, ``{Collapse of long-range charge order tracked by
  time-resolved photoemission at high momenta},'' \emph{Nature}, vol. 471, no.
  7339, pp. 490--494, 2011.

\bibitem{Hellmann2012}
S.~Hellmann, T.~Rohwer, M.~Kall{\"{a}}ne, K.~Hanff, C.~Sohrt, A.~Stange,
  A.~Carr, M.~M. Murnane, H.~C. Kapteyn, L.~Kipp, M.~Bauer, and K.~Rossnagel,
  ``{Time-domain classification of charge-density-wave insulators},''
  \emph{Nature Communications}, vol.~3, 2012.

\bibitem{Mathias2016}
S.~Mathias, S.~Eich, J.~Urbancic, S.~Michael, A.~V. Carr, S.~Emmerich,
  A.~Stange, T.~Popmintchev, T.~Rohwer, M.~Wiesenmayer, A.~Ruffing, S.~Jakobs,
  S.~Hellmann, P.~Matyba, C.~Chen, L.~Kipp, M.~Bauer, H.~C. Kapteyn, H.~C.
  Schneider, K.~Rossnagel, M.~M. Murnane, and M.~Aeschlimann, ``{Self-amplified
  photo-induced gap quenching in a correlated electron material},''
  \emph{Nature Communications}, vol.~7, pp. 1--8, 2016.

\bibitem{Smallwood2012x}
C.~L. Smallwood, J.~P. Hinton, C.~Jozwiak, W.~Zhang, J.~D. Koralek, H.~Eisaki,
  D.~H. Lee, J.~Orenstein, and A.~Lanzara, ``{Tracking Cooper pairs in a
  cuprate superconductor by ultrafast angle-resolved photoemission},''
  \emph{Science}, vol. 336, no. 6085, pp. 1137--1139, jun 2012.

\bibitem{HuberX}
M.~Huber, Y.~Lin, N.~Dale, R.~Sailus, S.~Tongay, R.~Kaindl, and A.~Lanzara,
  ``{Revealing the order parameter dynamics of 1T-\tise following optical
  excitation, submitted}.''

\bibitem{Buss2019}
J.~H. Buss, H.~Wang, Y.~Xu, J.~Maklar, F.~Joucken, L.~Zeng, S.~Stoll,
  C.~Jozwiak, J.~Pepper, Y.-d. Chuang, J.~D. Denlinger, Z.~Hussain, A.~Lanzara,
  and R.~A. Kaindl, ``{A setup for extreme-ultraviolet ultrafast angle-resolved
  photoelectron spectroscopy at 50-kHz repetition rate},'' vol.~90, no. 023105,
  2019.

\bibitem{Monney2009}
C.~Monney, H.~Cercellier, F.~Clerc, C.~Battaglia, E.~F. Schwier, C.~Didiot,
  M.~G. Garnier, H.~Beck, P.~Aebi, H.~Berger, L.~Forr{\'{o}}, and L.~Patthey,
  ``{Spontaneous exciton condensation in 1T$-$TiSe$_2$: BCS-like approach},''
  \emph{Physical Review B - Condensed Matter and Materials Physics}, vol.~79,
  no.~4, pp. 1--11, 2009.

\bibitem{DiSalvo1976}
F.~J. {Di Salvo}, D.~E. Moncton, and J.~V. Waszczak, ``{Electronic properties
  and superlattice formation in the semimetal TiSe$_2$},'' \emph{Physical
  Review B}, vol.~14, no.~10, pp. 4321--4328, nov 1976.

\bibitem{Kidd2002}
T.~E. Kidd, T.~Miller, M.~Y. Chou, and T.~C. Chiang, ``{Electron-hole coupling
  and the charge density wave transition in TiSe$_2$},'' \emph{Physical Review
  Letters}, vol.~88, no.~22, pp. 226\,402/1--226\,402/4, 2002.

\bibitem{Watson2019}
M.~D. Watson, O.~J. Clark, F.~Mazzola, I.~Markovi{\'{c}}, V.~Sunko, T.~K. Kim,
  K.~Rossnagel, and P.~D. King, ``{Orbital- and k$_z$-Selective Hybridization
  of Se$_{4p}$ and Ti$_{3d}$ States in the Charge Density Wave Phase of
  TiSe$_2$},'' \emph{Physical Review Letters}, vol. 122, no.~7, pp. 1--6, 2019.

\bibitem{Hellgren2017}
M.~Hellgren, J.~Baima, R.~Bianco, M.~Calandra, F.~Mauri, and L.~Wirtz,
  ``{Critical Role of the Exchange Interaction for the Electronic Structure and
  Charge-Density-Wave Formation in TiSe$_2$},'' \emph{Physical Review Letters},
  vol. 119, no.~17, pp. 1--6, 2017.

\bibitem{Chen2018}
C.~Chen, B.~Singh, H.~Lin, and V.~M. Pereira, ``{Reproduction of the Charge
  Density Wave Phase Diagram in 1T$-$TiSe$_2$ Exposes its Excitonic
  Character},'' \emph{Physical Review Letters}, vol. 121, no.~22, p. 226602,
  2018.

\bibitem{Bianco2015}
R.~Bianco, M.~Calandra, and F.~Mauri, ``{Electronic and vibrational properties
  of TiSe$_2$ in the charge-density-wave phase from first principles},''
  \emph{Physical Review B - Condensed Matter and Materials Physics}, vol.~92,
  no.~9, pp. 1--19, 2015.

\bibitem{Monney2010a}
C.~Monney, E.~F. Schwier, M.~G. Garnier, N.~Mariotti, C.~Didiot, H.~Beck,
  P.~Aebi, H.~Cercellier, J.~Marcus, C.~Battaglia, H.~Berger, and A.~N. Titov,
  ``{Temperature-dependent photoemission on 1T -TiSe$_2$: Interpretation within
  the exciton condensate phase model},'' \emph{Physical Review B - Condensed
  Matter and Materials Physics}, vol.~81, no.~15, pp. 1--9, 2010.

\bibitem{Cazzaniga2012}
M.~Cazzaniga, H.~Cercellier, M.~Holzmann, C.~Monney, P.~Aebi, G.~Onida, and
  V.~Olevano, ``{Ab initio many-body effects in TiSe$_2$: A possible excitonic
  insulator scenario from GW band-shape renormalization},'' \emph{Physical
  Review B - Condensed Matter and Materials Physics}, vol.~85, no.~19, pp.
  1--6, 2012.

\bibitem{Kohn1967}
W.~Kohn, ``{Excitonic phases},'' \emph{Physical Review Letters}, vol.~19,
  no.~8, pp. 439--442, aug 1967.

\bibitem{Rohde2014}
G.~Rohde, T.~Rohwer, A.~Stange, C.~Sohrt, K.~Hanff, L.~X. Yang, L.~Kipp,
  K.~Rossnagel, and M.~Bauer, ``{Does the excitation wavelength affect the
  ultrafast quenching dynamics of the charge-density wave in 1T$-$TiSe$_2$?}''
  \emph{Journal of Electron Spectroscopy and Related Phenomena}, vol. 195, pp.
  244--248, 2014.

\bibitem{Monney2016}
C.~Monney, M.~Puppin, C.~W. Nicholson, M.~Hoesch, R.~T. Chapman, E.~Springate,
  H.~Berger, A.~Magrez, C.~Cacho, R.~Ernstorfer, and M.~Wolf, ``{Revealing the
  role of electrons and phonons in the ultrafast recovery of charge density
  wave correlations in 1T$-$ TiSe$_2$},'' \emph{Physical Review B}, vol.~94,
  no.~16, pp. 1--9, 2016.

\bibitem{Snow2003}
C.~S. Snow, J.~F. Karpus, S.~L. Cooper, T.~E. Kidd, and T.~C. Chiang,
  ``{Quantum Melting of the Charge-Density-Wave State in 1T$-$TiSe$_2$},''
  \emph{Physical Review Letters}, vol.~91, no.~13, pp.
  1\,364\,021--1\,364\,024, 2003.

\bibitem{VanWezel2010}
J.~{Van Wezel}, P.~Nahai-Williamson, and S.~S. Saxena, ``{Exciton-phonon-driven
  charge density wave in TiSe$_2$},'' \emph{Physical Review B - Condensed
  Matter and Materials Physics}, vol.~81, no.~16, pp. 1--8, 2010.

\bibitem{Kaneko2018}
T.~Kaneko, Y.~Ohta, and S.~Yunoki, ``{Exciton-phonon cooperative mechanism of
  the triple- q charge-density-wave and antiferroelectric electron polarization
  in TiSe$_2$},'' \emph{Physical Review B}, vol.~97, no.~15, p. 155131, 2018.

\bibitem{Lian2020}
C.~Lian, S.~J. Zhang, S.~Q. Hu, M.~X. Guan, and S.~Meng, ``{Ultrafast charge
  ordering by self-amplified exciton–phonon dynamics in TiSe$_2$},''
  \emph{Nature Communications}, vol.~11, no.~1, 2020.

\end{thebibliography}

\section*{Acknowledgments}
This work was primarily funded by the U.S. Department of Energy (DOE), Office of Science, Office of Basic Energy Sciences, Materials Sciences and Engineering Division under contract no. DE-AC02-05CH11231 (Ultrafast Materials Science program KC2203). S.T.\ Acknowledges support from NSF-DMR 2111812 and NSF CMMI 1933214 for material development and characterization.

\section*{Competing Interest}
The authors declare no competing financial interest.

\section*{Author Contributions}
A.L.\ designed and supervised the project. Data was collected by M.H.\ and Y.L.\ and analyzed by M.H.\ with help from Y.L., N.D., R.K.\ and A.L.\ Samples were provided by R.S.\ and S.T.\ The XUV-trARPES setup was designed by R.K.\ Manuscript preparation was done by M.H.\ with input from all co-authors.



\end{document}